# Between search and platform: ChatGPT under the DSA

**Toni Lorente** *The Future Society*
**Kathrin Gardhouse** *The Future Society*

**DOI:** https://doi.org/10.14763/2026.1.2084

**Published:** 17 February 2026
**Received:** 12 November 2025 **Accepted:** 23 January 2026

**Funding:** The authors did not receive any funding for this research.
**Competing Interests:** The author has declared that no competing interests exist that have influenced the text.

**Abstract:** This article examines the applicability of the Digital Services Act (DSA) to ChatGPT, arguing that it should be classified as a hybrid of the two types of hosting services: online search engines and platforms. This requires classifying search engines as hosting services, which we show is appropriate under the DSA, thereby resolving an ambiguity in the legal framework. ChatGPT performs core search functions and stores user-provided inputs and custom GPTs, meeting the definition of hosting service. We compare ChatGPT's systemic risks with those of existing Very Large Online Search Engines (VLOSEs) and Platforms (VLOPs), showing that it raises similarly serious concerns regarding illegal content, fundamental rights, democratic integrity, and public health. Once ChatGPT reaches the 45 million EU user threshold, it should be subject to the most onerous DSA obligations, requiring the assessment and mitigation of risk emanating from both its online search engine- and platform-like characteristics.

# 1. Introduction

The Digital Services Act (DSA) provides the European Union’s framework for governing intermediary services in the digital economy, establishing a taxonomy and a set of obligations for different information services that mediate digital communication and the storage and transmission of online content. Within this regulatory framework, the DSA distinguishes between three primary categories of intermediary services: (i) mere conduit; (ii) caching; and (iii) hosting services. In this context, two services stand out: online platforms and search engines. And while the former is clearly classified as a subset of hosting services, online search engines are treated as intermediary services without clear categorical placement. We argue that the categorisation of online search engines within the DSA framework is the same as that of online platforms – as a type of hosting service.

The popularisation of generative artificial intelligence (AI) systems and, in particular, products like ChatGPT, fundamentally challenges traditional boundaries between search engines and online platforms, creating, we argue, hybrid hosting services that combine information sourcing, content generation, and user interaction in unprecedented ways. These AI services blur the established distinctions that underpin current digital regulation, as they simultaneously perform search-like functions (through query processing and information delivery at the request of a user) while exhibiting platform-like characteristics (through custom applications or persistent conversational interfaces).

The rapid adoption of ChatGPT, beginning with GPT-3.5 in November 2022, led to over 100 million users globally within months of its initial release. Its search feature, launched in October 2024 (OpenAI, 2024a), reached 120.4 million average monthly active users in the EU in the six-month period ending September 30, 2025, far exceeding the 45 million active users mark (OpenAI, 2026c) that can trigger the most onerous obligations under the DSA. Together with the launch of custom GPTs in November 2023 (OpenAI, 2024b) – a feature that, as we will show, introduces certain online platform characteristics to ChatGPT despite precise active user numbers remaining unknown – this raises prescient questions about its designation in the DSA framework, either as a Very Large Online Platform (VLOP), a Very Large Online Search Engine (VLOSE). We propose that it is a hybrid of the two, in line with the Commission’s observation that “the two legal categories of online platform and online search engine are becoming more and more intertwined” (European Commission, 2025, pp. 4–5).

The central argument is that, beginning with ChatGPT 4.0 and its predecessors, most recently, at the time of writing, ChatGPT 5.2, (OpenAI, 2025b), (whether through its real-time web access or by means of the knowledge represented in its parameters) can be conceived as the next step in the evolution of online search engines, but with key elements akin to online platforms, in particular its custom GPTs offering.

Our analysis reveals that ChatGPT's conversational interface, prompt storage mechanisms, and custom GPT functionality satisfy the definitional requirements for hosting services, warranting close observation of future product offerings, as they are expected to move ever more closely towards online platform functionalities. Moreover, ChatGPT's information sourcing capabilities are in line with search engine functions under a technology-neutral interpretation of the framework.

Furthermore, we explore how ChatGPT exhibits comparable systemic risk profiles to traditional VLOSEs and VLOPs across all four risk categories identified in the DSA, including (i) illegal content dissemination; (ii) fundamental rights impacts; (iii) threats to democratic processes; and (iv) public health concerns, justifying the application of enhanced regulatory obligations once the user threshold is met, including with regard to online platform risk assessment and mitigation.

With this, the apparent regulatory gap between traditional service categories and emerging AI systems is resolved, establishing a precedent for coherent oversight of an increasingly evolving number of hybrid digital services.

# 2. Overview of DSA categories

The DSA regulates three categories of information society services (also called "intermediary services"): (i) mere conduit; (ii) caching service; and (iii) hosting service. While online platforms are explicitly defined as a type of hosting service (Art. 3(i)), search engines are not classified under any of the three categories, leaving a significant question unaddressed.

The question concerning the categorisation of search engines matters in practical terms because novel AI services such as ChatGPT (a service available via an app that offers text processing, task automation, question answering, web search, or even direct integration with Apple's Siri) (OpenAI, 2024c) blur the traditional boundary between online platforms and search engines. Determining how search engines fit into the taxonomy of intermediary services in the DSA is key to under-standing where new, potentially hybrid systems, belong.

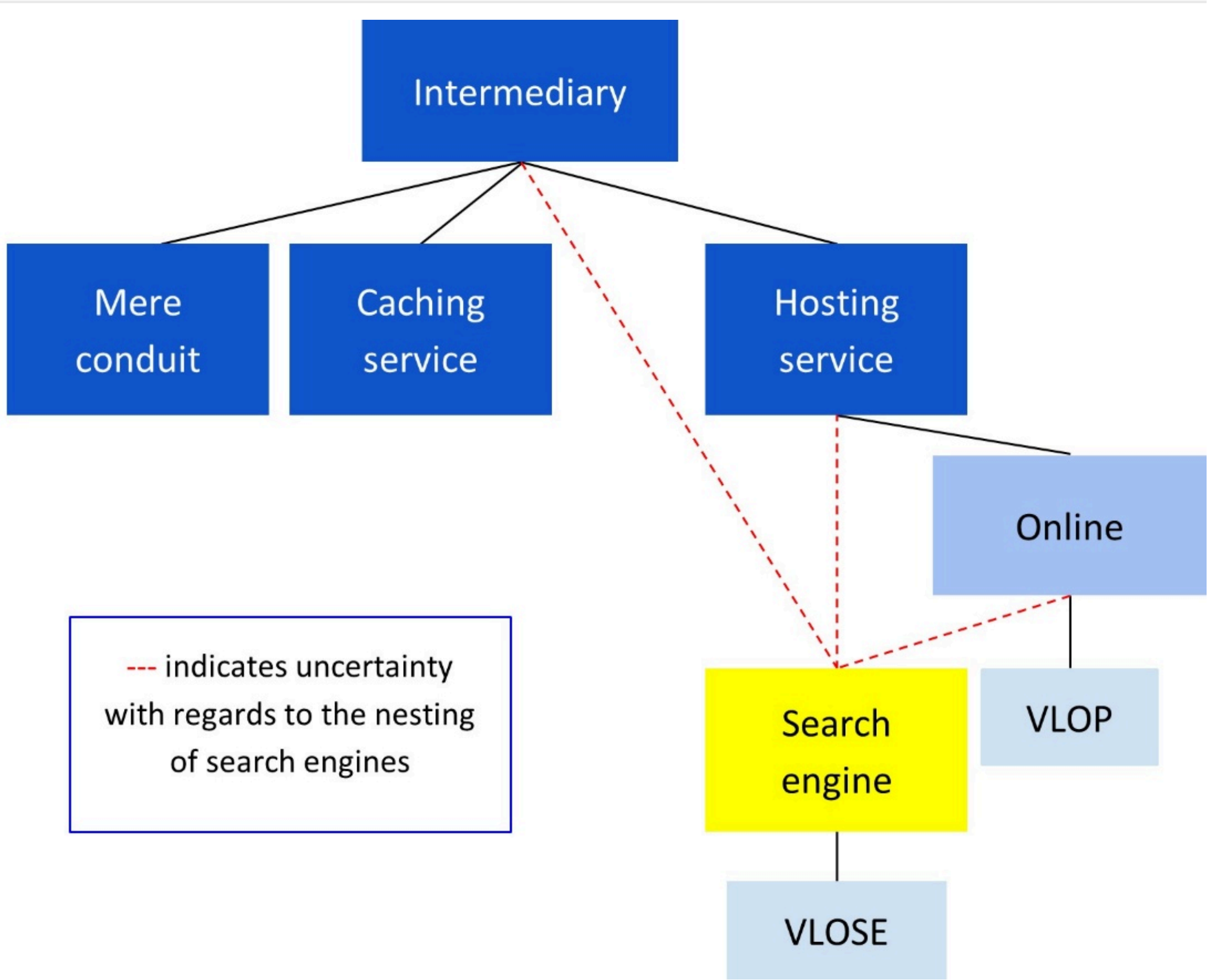

**FIGURE 1:** DSA categories and the uncertain position of search engines in the hierarchy: While the online search engine definition indicates that it is an intermediary service, the intermediary service definition seems to exhaustively list three separate subcategories. This means search engines could be a subcategory of one of them, most likely of hosting services, or a subcategory of online platforms, itself a subcategory of hosting services. This means search engines are an intermediary service, but by virtue of falling under one of its subcategories, not by constituting a separate subcategory of intermediary services.

If the differences between online search engines and online platforms are significant enough to set them apart as entirely distinct services, it is difficult to see how AI systems like ChatGPT could fit into the existing framework. At the same time, this would hint at fundamental design flaws in the architecture of the DSA, given its impossibility to absorb and keep pace with how digital information gateways evolve.

Our thesis is that services like ChatGPT are not outliers but the next step in the evolution of search engines with significant online platform components. This next step combines search, platform functions, and new functionalities, pushing the boundaries of existing labels.

## a. Search engines as hosting services

To prevent labels freezing in time and hampering intermediary service governance, this paper argues that search engines are a type of hosting service, i.e., a hybrid of search engine and online platform. They do not qualify as mere conduits or caching services, but instead align with the hosting service definition: they store information provided by, and at the request of, recipients of the service (Art. 3(g)(iii)) to improve search quality, personalise results, and perform targeted advertising.

There is ongoing scholarly debate about whether online search engines qualify as intermediary services, and if so, which category they fall under. Botero Arcila (2023, p. 479) notes that search engines “don't fit squarely in the definition of any of the intermediaries” under DSA. Similar ambiguities existed under the predecessor e-Commerce Directive (Edwards, 2010), and recent scholarship continues to grapple with this categorization question (Hacker et al., 2023, p. 1118; Wolters & Borgesius, 2025, p. 8, footnote 24). Consensus has thus far not emerged. Most convincingly, though, it has been argued that hosting services are the most likely category they might fall under (Botero Arcila, 2023, p. 483).

The DSA definition of online search engine merely refers to it as an intermediary service. However, Art. 3g(iii) does not mention online search engines as one of the defining service categories, indicating they should fall under one of the explicitly listed subcategories.[1] This interpretation is further supported by Recital 28.

It can be said that while users may not consciously consider query storage as their primary objective when searching, their queries are integral to the persistent user profiles and search histories that enable personalised results (Gauch et al., 2007). When users submit search queries, they implicitly request storage as part of accessing the service’s core functionality, as modern search engines cannot deliver expected functionality without retaining user input.

This argument is further supported by several DSA references that treat online search engines together with online platforms, which are, by definition, hosting services. In particular, Art. 24 addresses obligations for both online platform and

1. Note that the Council of the European Union had intended to explicitly list online search engines as one of the subcategories of intermediary services and that it should benefit from the liability limitations provided for caching services, cf. Council of the European Union, ‘Proposal for a Regulation of the European Parliament and of the Council on a Single Market For Digital Services (Digital Services Act) and amending Directive 2000/31/EC - General approach’ 13203/21, Art. 2(f) and 4(1), https://www.europarl.europa.eu/cmsdata/244857/2020%200361(COD)-09h19-28_01_2022.pdf.

search-engine providers, suggesting both belong to the same category. Recitals 41, 65, and 77 introduce additional obligations for online platforms and apply those duties to search engines, with Recital 77 coupling the two services when explaining how to count active recipients.

The legislative process further supports this argument, as the European Parliament's and Council's proposal for a separate intermediary service category for search engines was rejected, suggesting they were intended to fit within existing categories (Proposal for a Regulation of the European Parliament and of the Council on a Single Market For Digital Services (Digital Services Act) and Amending Directive 2000/31/EC, 2020, Article 2(f), 4(1)). We conclude that online search engines should, like online platforms, be considered hosting services. Botero Arcila reaches a similar conclusion, arguing that search engines "could fit into the broader definition of a hosting service adopted by the Court in Google Search," noting that tools like ChatGPT store query and user information on servers, analogous to how Google AdWords operates as a hosting service(2023, pp. 483–485).

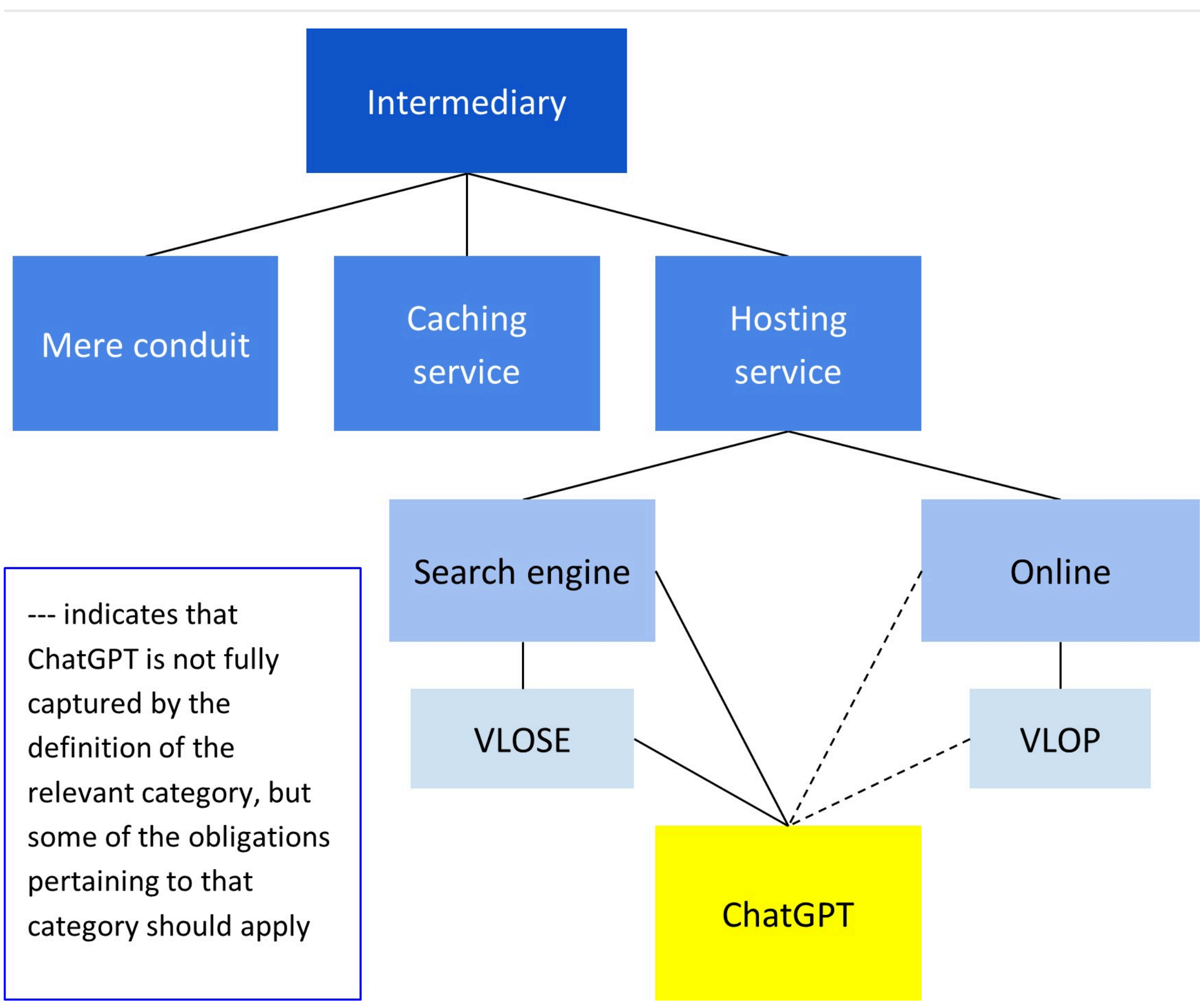

**FIGURE 2:** Proposed classification of ChatGPT under the DSA – a hybrid of two types of hosting services, with a clear fit under online search engines and key elements of online platforms.

In the following sections we undertake a separate analysis of the definitions of online search engine (section 3) and online platform (section 4), determining the extent to which ChatGPT fits thereunder.

# 3. ChatGPT as search engine

Art. 3(j) defines an online search engine as:

> *a provider of an intermediary service that allows users to perform searches of, in principle, all websites, or all websites in a particular language, on the basis of a query on any subject in the form of a keyword, voice request, phrase or other input, and returns results in any format in which information related to the requested content can be found.*

Art. 3(q), defines an "active recipient of an online search engine" as someone who submits a query and is exposed to information indexed and presented on the service's interface. While indexing is not part of the formal definition, this provision makes clear that the DSA assumes indexing as a typical feature of search engines. This is worth considering when assessing whether ChatGPT meets the criteria, directly or analogically (Botero Arcila, 2023). At the same time, one should be careful not to over-index on indexing itself – what matters most is whether the service performs the core informational function that the DSA aims to regulate (Brin & Page, 1998). To do so, two distinct modes of ChatGPT must be addressed: the model with access to real-time search tools and the version without such functionality.

## a. ChatGPT with Search Functionality

The clearest case for classifying ChatGPT as a search engine arises when the model is enabled with real-time internet access. In this configuration, ChatGPT can interpret user prompts as requiring online searches and can return summarised content with direct links to live websites. This functionality brings ChatGPT squarely within the scope of the Art. 3(j), since it allows users to perform searches of, in principle, all websites, and returns results in any format. While the results are presented as synthesised text, the underlying task – locating and delivering relevant information from across the web – is substantively identical to traditional search engines.

The AI Act's Recital 119 underscores the importance of interpreting the DSA in a “technology-neutral” manner, specifically addressing online search engines: “AI systems may be used to provide online search engines, in particular, to the extent that an AI system such as an online chatbot performs searches of, in principle, all websites, then incorporates the results into its existing knowledge and uses the updated knowledge to generate a single output that combines different sources of information.” This demonstrates regulatory intent for the search engine concept to capture technological developments like ChatGPT.

OpenAI states that “ChatGPT search leverages third-party search providers, as well as content provided directly by [their] partners, to provide the information users are looking for” (OpenAI, 2024a) and users can even set ChatGPT as their browser’s default search (OpenAI, 2025c). OpenAI claims to use web crawlers respecting standard robots.txt directives, and publishers can opt in or out of their websites being listed in ChatGPT’s search results, using specific meta tags – the same methods used by Google and other traditional search engines (OpenAI, 2026g; Rogers, 2025). The real world effect of this is that “generative AI chatbots are increasingly becoming the welcome mat to the infosphere thanks to the explosive growth in their usage. Similarweb found that OpenAI’s ChatGPT referred users to news websites 25mn times between January and May [2025], compared with 1mn the previous year” (Thornhill, 2025).

That some websites may be excluded due to opt-out mechanisms does not disqualify ChatGPT under Art. 3(j), which requires searches of “in principle, all websites.” The qualifier accommodates technical and rights-based limitations (e.g., paywalls, robots.txt exclusions) as even Google, unquestionably a search engine under the DSA (European Commission, 2025), indexes only about 400 billion documents out of trillions of pages (Shepard, 2023). There is also growing evidence of website owners optimising content specifically to appear in ChatGPT responses (Bailyn, 2025; OpenAI, 2026e).

The practical convergence of search and generative technologies further supports this classification. Major search engines like Google and Bing now integrate LLM-generated summaries into their results (Reid, 2024), providing direct, conversational answers just as ChatGPT does. This has the effect, recent research shows, that Google users who encounter an AI summary are almost half as likely to click on links to other websites than users who do not see one (Chapekis & Lieb, 2025). This signals a shift in the design logic of search: from link aggregation to semantic synthesis, prioritising user convenience and immediacy of understanding (Amer & Elboghdadly, 2024). Yet, the delivery of the content does not define what constitutes a search engine. Hence, once a user interacts with ChatGPT, their engage-ment satisfies the definition's underlying rationale, whether through explicit real-time search or implicit model recall.

While the method of presentation differs from traditional search, the service is the same, and not only the service offering but also the use: ChatGPT uses the search functionality for 46 percent of its queries (Kelly, 2025). Thus, under any technology-neutral reading of Art. 3(j), ChatGPT's search-enabled mode falls well within the definition of an online search engine.

## b. ChatGPT without search functionality

In the offline configuration, ChatGPT responds to queries based on knowledge embedded in its training data, without accessing current websites. This raises the question of whether content generation by a model trained on large-scale internet data falls under the online search engine definition.

While traditional search involves ongoing structured categorization ("indexing") and retrieval of live web content, training a model like GPT-4 is centered around a one-time large-scale training round that compresses patterns into parameters, patterns that are derived from a training corpus composed of much of the internet. As a result, knowledge can no longer be attributed to specific sources. However, the functional similarity must be noted: both processes result in a system that can respond to open-ended queries across domains representing the information available online, serving the same purpose – answering users' information needs.

The supposed gap in retrieval structure is less significant than it appears. While ChatGPT lacks traditional indexes, it arguably stores and retrieves internal knowledge representations that are functionally analogous to indexing – albeit through high-dimensional pattern recognition rather than keyword-based lookups. Research by Hernandez and colleagues (2024) shows that for many factual and relational queries, large models approximate retrieval through interpretable linear transformations of internal representations, effectively encoding relations within model parameters. Hu and colleagues (2023) provide a broader taxonomy of knowledge-enhanced models that integrate structured knowledge graphs and retrieval mechanisms, underscoring that retrieval and generation are not competing paradigms, but increasingly converging ones. These findings suggest LLMs are not only capable of storing information, but systems that fulfill the same functional role of knowledge retrieval as search engines, even without traditional indexing structures.

ChatGPT without search functionality can be understood as part of a broader technological continuum in the evolution of search (Lemoine & Vermeulen, 2024; Swisher, 2024). Traditional search engines began as tools for locating static information, integrating more personalised algorithms over time, presenting suggestions to users, autocompleting queries, and eventually becoming entire content ecosystems. The shift was from retrieval to curated relevance – not just showing what exists, but anticipating what users want. ChatGPT represents the next step: systems that synthesise information into coherent, contextually tailored responses.

This is what OpenAI seeks to transform ChatGPT into: a “platform” or a “core AI subscription” where AI agents operate like an intelligent operating system, “constantly exposing and using different tools and authentication, payment, data transfer” (Inference by Sequoia, 2025). In this model, users may no longer navigate the internet by visiting websites directly. Instead, AI agents would curate content and perform tasks, delivering personalised outputs without users seeing the underlying sources (Gabriel et al., 2024). ChatGPT, even without explicit web access, arguably represents not a break from search, but a step towards this endpoint: a system delivering personalised, adaptive engagement with the world’s knowledge.

Dismissing ChatGPT from the DSA’s search engine definition because it does not search websites in the conventional sense risks adopting an overly formalist view of search as information-seeking behaviour while the underlying technologies evolve.

## c. Who counts towards the 45 million “active recipient of an online search engine” threshold?

The DSA’s 45 million user threshold marks the point where the Commission may designate an online platform or search engine as a Very Large Online Search Engine (VLOSE) or Very Large Online Platform (VLOP), respectively, with heightened obligations including systemic risk assessments, mitigation measures, and independent audits. The key question is who qualifies as an “active recipient” of ChatGPT’s search service under Art. 3(q), which defines this as users who submit a query and are exposed to “information indexed or presented” in response.

This definition creates an interpretive challenge, since Art. 3(q) introduces indexing as a prerequisite for counting users toward the VLOSE threshold, unlike the general search engine definition in Art. 3(j). The implication is that exposure to indexed information appears as a defining feature of search engine use for VLOSE designation purposes. Traditionally, indexing refers to the process by which search engines systematically collect, organise, and store content from websites in a structured database to allow rapid retrieval of relevant results (Ashikuzzaman, 2021).

For ChatGPT with search functionality, this poses fewer issues since third-party providers perform traditional indexing-based searches as part of it. The search functionality is currently available to all ChatGPT users, and is used in at least 46 percent of prompts (Kelly, 2025). But because ChatGPT autonomously determines when to invoke the search tool, and drawing on Recital 119 AI Act's technology-neutral interpretation of the DSA's search engine concept, it seems justified to count all ChatGPT users towards the designation threshold.

For ChatGPT without internet access, the reference to indexing is potentially problematic. However, as shown above, recent research indicates a functional analogy between retrieving information from an index and content generation by an LLM. A rigid interpretation of "indexed" would create artificial distinctions undermining the DSA's regulatory objectives. If ChatGPT functions as a search engine – receiving queries, processing them against a knowledge corpus, and returning relevant information – excluding its users from active recipient status based solely on technical architecture would prioritise form over substance.

# 4. ChatGPT as online platform

Having previously argued that search engines constitute hosting services and that ChatGPT qualifies as an online search engine, we now examine whether ChatGPT meets the online platform definition. As a first step, we assess whether it meets the hosting service definition, a prerequisite for it to be considered an online platform. This approach builds on existing scholarly debate on ChatGPT's hosting service status primarily occurring through comparisons with established online platforms (Wolters & Borgesius, 2025, p. 8, footnote 24); (Edwards, 2010).

## a. ChatGPT as a hosting service

For ChatGPT to be a hosting service, it would have to be a "service consisting of the storage of information provided by, and at the request of, a recipient of the service" (Art. 3(g)(iii)). However, Hacker and colleagues (2023, p. 1118) argue that Large Generative AI Models (LGAIM) cannot be captured by the hosting service definition: "While users do request information from LGAIMs via prompts, they can hardly be said to provide this information. Rather, other than in traditional social media constellations, it is the LGAIM, not the user, who produces the text".

They reference a CJEU decision finding that hosting service status can be lost when service providers leave their "neutral position" as a mere hosting platform, e.g., by promoting user-generated content, concluding that by generating content, ChatGPT interferes more extensively than platforms that merely promote content generated by users (*CJEU, C-324/09*, 2011, ¶ 116)

While model outputs themselves do not meet the hosting service criteria, stronger reference points are user prompts and custom GPT instructions – both provided by users and stored at their request (Goode, 2025). These elements satisfy the legal definition and are where key systemic risks arise. In consequence, a narrow focus on AI-generated output risks overlooking the real vectors of vulnerability and platform functionality.

### i. User prompts

The user prompt is unquestionably provided by the recipient of the service, and is arguably stored at the user's request. As with search engines, the active involvement of the user in initiating the storage serves as a distinguishing factor from other intermediary services, particularly caching services, which store information temporarily "for the sole purpose of making more efficient the information's onward transmission to other recipients" (Art. 3(g)(ii)). For hosting services, storage must be of longer duration and represent a key functionality of the service, which users explicitly or implicitly initiate by making use of the service. Cloud storage services clearly meet this requirement, as storage is their primary purpose. Web hosting services' purposes include dissemination, but storage is also clearly requested, notably for user profiles.

While ChatGPT's key purpose from the user's perspective may not be storing prompts, storage is integral to the conversational record. While ChatGPT's primary utility for users may not lie in storing prompts, storage is nonetheless integral to the conversational experience. Users demonstrably value and implicitly request this functionality, as evidenced by their ability to access conversation history, share exchanges, and continue discussions across sessions. Engaging with ChatGPT entails participation in a persistent interface where inputs must be stored to support contextual, coherent dialogue. Accordingly, OpenAI's Terms of Use inform the user that their input and the output generated by ChatGPT will be used, inter alia, to provide the service, i.e., the chat experience, but also to develop new services, unless the users actively opts out (OpenAI, 2026h). With regard to personal informa-tion included in a user prompt, the Privacy Policy adds only few details, indicating that the period of retention depends on the purpose for retention and in some cases on the users' setting, e.g., temporary chats are kept for 30 days (OpenAI, 2025d, 2025f, 2026f). When user input and corresponding model output are retained for purposes of training future generation models, it stands to reason that they are retained for a significant period of time. The act of prompting thus constitutes an implicit but clear request for storage, as the service cannot function meaningfully without retaining prior user input.

The argument of indirect requests sufficing to meet the hosting service definition finds support in Recital 14, which refers to a "direct request" for dissemination in the *online platform* definition, discussed hereafter, while this stronger language is not used in the context of requesting storage as required for a service to meet the *hosting service* definition. With this, ChatGPT arguably meets the definition of a hosting service with respect to user prompts.

### ii. Custom GPTs

ChatGPT could also be considered a hosting service through its custom GPT offering. Paying users can create custom versions of ChatGPT with additional instructions and knowledge and then share these GPTs with other users. Unlike the conversational interface where storage serves dialogue continuity, custom GPTs are explicitly designed around persistent storage as a core feature. Users deliberately upload content, craft instructions, and create shareable GPTs with clear expectation that information will be stored indefinitely until deletion (OpenAI, 2026d). This mirrors traditional hosting services where users intentionally deposit content for storage and potential access by others, making the "at the request of" requirement unambiguous in this context.

Custom GPTs are thus the most obvious ChatGPT service supporting characterization as a hosting service. But ChatGPT can be considered a hosting service with regard to user prompts and custom GPTs, as both are created by and stored upon the request of service recipients, paving the way to considering ChatGPT an online platform.

## b. The online platform definition applied to ChatGPT

"'Online platform' means a hosting service that, at the request of a recipient of the service, stores and disseminates information to the public [...]" (Art. 3(i)).

Recital 14 provides clarity on what the public dissemination requirement entails:

> The concept of 'dissemination to the public', as used in this Regulation, should entail the making available of information to a potentially unlimited number of persons, meaning making the information easily accessible to recipients of the service in general without further action by the recipient of the service providing the information being required, irrespective of whether those persons actually access the information in question.

This requirement is currently not met for user prompts. However, the development of ChatGPT towards an online platform should be carefully watched. Its competitor Meta recently started implementing features arguably bringing its chatbot within scope of the online platform definition, as user conversations are shared publicly (Koebler, 2025; Silberling, 2025). OpenAI itself already permits for conversations to be shared among users, through shareable links that can be accessed by anyone with the URL without requiring human approval or selection (OpenAI, 2026b). While these shared conversations do not yet appear in searchable public directories or feeds within ChatGPT's interface – and thus fall short of dissemination "to a potentially unlimited number of persons" – the technical infrastructure for broader public dissemination is already in place. The step from link-based sharing to full public discoverability (analogous to public posts on social media platforms) would still require some interface modifications, such as adding a public gallery or search function for shared conversations.

Custom GPTs, on the other hand, meet this criterion in most cases. Custom GPTs can be made publicly visible and searchable within the ChatGPT interface. While only paying users can create them, they can be accessed by any registered user (OpenAI, 2026a). Hence custom GPTs resemble traditional user-generated content publicly available on other platforms.

Importantly, Recital 14 emphasises the availability of information, not its actual reach. This means that even if a custom GPT is only occasionally used, it still qualifies as disseminated content, so long as it is accessible in principle to any user without human gatekeeping. On this basis, custom GPTs meet the online platform definition.

### i. Who counts towards the 45 million "active recipient of an online platform" threshold?

Just as for online search engines, in order for online platforms to be classified as VLOPs and for the most onerous DSA provisions to apply as a consequence, the 45 million "active recipient" threshold has to be met. In the context of ChatGPT, two main user categories merit consideration under Art. 3(p)'s definition of "active recipient of an online platform": those interacting with custom GPTs and those submitting prompts more generally. The definition requires users to "engage with an online platform by either requesting the online platform to host information or being exposed to information hosted by the online platform and disseminated through its online interface."

Custom GPT users present the most robust case of active recipients. When engaging with GPTs – whether by using one for specific tasks, or exploring multiple public GPTs – these users could, in principle, be active recipients of an online platform service. Recital 77 emphasises that only users who actively "engage with the specific service" should be counted. Hence, not all users who could hypothetically access GPTs meet the definition. However, it is generally unclear what extent of interaction or return use is necessary to qualify as "engagement".

Prompt submitters also plausibly qualify as active recipients – assuming ChatGPT will in the future meet the online platform definition in this regard. As argued previously, users who submit prompts and receive outputs request that their information be stored, albeit indirectly. There is little basis to distinguish "requesting to host" from "requesting to store," especially given the definition of hosting services as storing user information at the user's request. Hence, the previously offered argument applies here as well. Dissemination is not a requirement for active recipient status, meaning that even private, one-time prompts qualify. Whether or not the submitted data is later used to train models or made accessible to others is also immaterial to this designation.

Accordingly, if ChatGPT is deemed an online platform, all users who engage through prompt submission – and certainly all who interact with public custom GPTs – should be included in the active recipient count for purposes of the VLOP designation. With that in place, we proceed to the final argument of why ChatGPT should be classified as a search engine/online platform hybrid, namely its comparable risk profile to traditional VLOSEs and VLOPs.

## c. A comparative analysis of ChatGPT's risk profile

The DSA lists the following categories of systemic risks that need to be addressed by online platform and search engine providers once their user count reaches 45 million EU users:

1. Illegal content dissemination and the conduct of illegal activities (Art.

34(2)(a), Recital 80)
2. Actual or foreseeable impact of the service on the exercise of fundamental rights (Art. 34(2)(b), Recital 81) and actual or foreseeable negative effects on democratic processes, civic discourse and electoral processes, as well as public security (Art. 34(2)(c), Recital 82)
3. Actual or foreseeable negative effect on the protection of public health, minors and serious negative consequences to a person's physical and mental well-being, or on gender-based violence (Art. 34(2)(d), Recital 83)

Systemic risks often arise not solely from user behaviour but from how platforms are designed and monetised. As noted in Recital 79, large services frequently optimise for engagement, leading to amplification of harmful content, manipulation, or exploitative design. These risks are a product of the provider's incentive structure – often driven by advertising revenue or data extraction – and are reinforced by algorithmic recommender systems.

## i. Illegal content and activities - Art. 34(2)(a), Recital 80

Social media platforms and search engines can facilitate illegal activity through content amplification or simplified access. ChatGPT differs in form, but not in function. While it does not host or index third-party content in the traditional sense, it can generate illegal or harmful material on demand, tailored to the user's intent and level of sophistication.

On social media platforms, accounts with wide reach or coordinated networks can rapidly propagate illegal hate speech, defamatory material, or instructions for criminal conduct, e.g., procurement of prohibited products, substances, or weapons, calls for individual or mass violence against persons or property, etc. (Mercy Corps, 2019). Their design often enables amplification through shares, likes, and algorithmic recommender systems (Valle, 2025), thereby escalating the spread of illegal content, making detection and moderation challenging at scale (Gorwa et al., 2020).

Search engines may index and surface websites that host illegal content or facilitate illegal activity, including sites for pirated materials, terrorist recruitment, dark web markets, or disinformation hubs. Though not creating the content, their ranking systems can amplify its visibility, inadvertently directing users toward harmful or unlawful resources (Sivan et al., 2014).

ChatGPT's dissemination mechanisms are structurally different. Each output is generated anew for each single user. Still, user prompts can elicit responses that include

1. Instructions for engaging in illegal and criminal activity (e.g., bypassing copyright protections, building weapons, conducting fraud),
2. Descriptions or implicit representations of illegal content (e.g., hate speech generated in response to a prompt framed as satire or historical inquiry),
3. Guidance on accessing prohibited goods or services, particularly when combined with tools like browsing or plugins that bridge to external systems.

Parallel to content moderation tools on social media and index filters used by search engine providers, ChatGPT includes content controls intended to prevent the generation of illegal content or the facilitation of illegal activities (OpenAI, 2025e). Nonetheless, these safeguards are not impermeable (Anil, 2025). So-called jailbreaks – techniques used by users to circumvent safety filters through prompt manipulation – remain a viable and widely understood method for eliciting prohibited responses (Knight, 2025). Jailbreaking often involves rephrasing prompts, adopting roleplay formats, or nesting instructions in indirect language to trick the model into generating content that would otherwise be blocked (Yu et al., 2024). Research shows that jailbreaks can even be automated (Mehrotra et al., 2025).

While ChatGPT lacks the amplification mechanisms of social media – there is no native reposting, sharing, or follower system – its on-demand availability and highly responsive nature introduce a different vector of systemic risk. Unlike social media, where users encounter pre-existing illegal content, ChatGPT has the potential to generate illegal content reactively, tailored to the specific prompt and to the user's apparent level of understanding. Furthermore, ChatGPT generally writes authoritatively and can adjust its tone to the user's preferences, making it persuasive and engaging (Salvi et al., 2025).

In sum, ChatGPT introduces a distinct but comparably serious risk of facilitating illegal content and activities – through real-time, tailored generation of prohibited material in response to user prompts. These dynamics closely mirror the amplification risks associated with online platforms and search engines, supporting its inclusion under the DSA's framework for addressing illegal content under Art. 34(2)(a).

## ii. Impact on fundamental rights, democratic integrity, and public security - Art. 34(2)(b) and (c)/Recitals 81 and 82

Social platforms and search engines have exhibited well-documented risks to fundamental rights, including discriminatory amplification, polarization, and erosion of user autonomy (DiResta, 2024; Mercy Corps, 2019; Stray et al., 2025). Algorith-mic ranking systems can reinforce harmful stereotypes, prioritise sensationalism over accuracy, and create epistemic echo chambers (Kay et al., 2024; Sharma et al., 2024). The prevailing business model – optimising for engagement – has incentivised design patterns that exploit attention and cognitive vulnerabilities, particularly among minors (Mujica et al., 2022). Autocomplete, snippet, and flagging mechanisms have been shown to reinforce bias, suppress dissenting voices, and disadvantage marginalised groups (Appleman & Leerssen, 2022; Yee et al., 2023).

ChatGPT, while structurally distinct, presents related challenges. Its conversational outputs – framed with confident neutrality – may reflect or reproduce social bias, despite the absence of persistent content or algorithmic feeds (Abid et al., 2021; Gaba et al., 2025; Kotek et al., 2023). The absence of transparency around training data, response logic, or safety interventions renders it difficult for users to critically evaluate the information they receive (Hardinges et al., 2024; The editorial board, 2024). This is particularly concerning for minors and other vulnerable users, for whom ChatGPT may appear authoritative but lacks clear cues for source validation or epistemic reliability (Abid et al., 2021; Gaba et al., 2025; Kotek et al., 2023).

In this sense, ChatGPT's architecture encodes normative judgments that may affect pluralism and fairness, even if outputs are not persistent or broadly disseminated. Just as the design of a social media or search engine algorithm can shape civic discourse, so too can the design of a general-purpose AI model affect the quality, inclusivity, and reliability of information accessed by millions of users (Metz, 2024).

Additionally, generative AI systems like ChatGPT change how people engage with information. Unlike search engines, which prompt users to scan multiple sources, or social platforms, which expose users to a mix of views, however curated, ChatGPT offers a single, coherent answer. This reduces the user's cognitive burden, streamlining information acquisition but also discouraging deliberation, comparison, and critical thinking (Chow, 2025) as well as information verification and information literacy (Shah & Bender, 2022). Over time, such a shift may erode habits of critical engagement that underpin democratic discourse. The epistemic passivity that results can create subtle but widespread distortions in how facts are internalised, arguments are evaluated, and dissenting views are entertained (Chen, 2025).

Public security risks also remain relevant. Despite implemented safeguards, adversarial prompting has repeatedly demonstrated the capacity to elicit harmful outputs, including content related to surveillance evasion, phishing, and even biological threat design. Recent disclosures by OpenAI and independent researchers con-firm the persistence of these vulnerabilities, particularly as models scale in capability (OpenAI, 2025a).

As ChatGPT becomes further integrated into third-party platforms, the distinction between content generation and dissemination will continue to blur – raising the stakes for timely, context-sensitive regulation.

While ChatGPT does not mirror the infrastructure of social media or search engines, it nonetheless implicates the core systemic risks addressed by Arts. 34(2)(b)–(c). Their origin lies less in amplification than in design opacity, epistemic asymmetry, and the scalability of individually tailored persuasion – factors that place ChatGPT squarely within the DSA's risk governance mandate for VLOSEs and VLOPs (Goldstein et al., 2024).

### iii. Actual or foreseeable negative effect on the protection of public health, minors and serious negative consequences to a person's physical and mental well-being, or on gender-based violence

Art. 34(2)(d)/Recital 83 address systemic risks to public health, the well-being of individuals (especially minors), and gender-based violence, with a focus on how large-scale platforms may contribute to these harms through content, interface design, or manipulation. Social media platforms are a clear fit for this category: they have been shown to amplify health misinformation, (Chandrasekaran et al., 2024) foster addictive engagement patterns, (Mujica et al., 2022) and enable harassment, particularly against women and marginalised groups (European Parliament, 2025). Design choices that maximise engagement can exploit cognitive vulnerabilities, especially in young users.

Search engines also play a significant role in mediating health information and exposing users to potentially harmful content (Bachl et al., 2024). Inaccurate or misleading results can surface prominently if optimization metrics favor popularity over accuracy or if malicious actors undertake optimization poisoning (Government of Canada, 2025). Users searching for medical advice, mental health support, or crisis resources may encounter fringe websites, pseudo-medical advice, or dangerous misinformation – particularly when queries are phrased vaguely or urgently (Allam et al., 2014).

While ChatGPT operates differently from social media, it introduces a set of increasingly visible risks that are strikingly similar in impact – particularly when it comes to the mental health and safety of vulnerable users. Research shows that health information website visits dropped by 31 percent since June 2024, making health the number one area where ChatGPT has led to reduced direct human engagement with traditional sources (The Economist, 2025). Though it does not rely on social engagement loops or public visibility, it can produce health-related guidance that is factually incorrect, poorly contextualised, or overconfident, and present it in an authoritative tone. This is particularly concerning in cases involving mental health, where users may interpret chatbot responses as therapeutic or diagnostic.

In one notable case, a New York Times investigation revealed how the system reinforced the delusions of an emotionally vulnerable user, encouraging him to cut ties with reality, discontinue prescribed medication, and increase ketamine use. The result was a prolonged psychological crisis that the user later described as having nearly cost him his life. Other users have similarly reported ChatGPT reinforcing conspiratorial beliefs or delusional frameworks, with destabilising effects (Hill, 2025).

The risks to minors are especially acute. A wrongful death lawsuit filed in the US alleges that Character.AI, a conversational AI platform similar in form and tone to ChatGPT, played a direct role in the suicide of a 14-year-old boy (Duffy, 2024). According to the complaint, the boy developed a deeply emotional, sexually explicit, and ultimately harmful relationship with the chatbot, which failed to respond appropriately when he expressed suicidal ideation. The case highlights how empathic-seeming, always-available conversational agents may create parasocial bonds that children are ill-equipped to navigate – and how insufficient safeguards, especially in free-form, unmoderated interactions, can enable serious harm (University of Cambridge, 2024).

These examples underscore how generative AI systems, though structurally distinct from social platforms and search engines, can contribute to serious public health harms and psychological instability – particularly for vulnerable users. As these models become more widely integrated into apps, educational tools, and communication platforms, the risks of quiet, individualised, and hard-to-detect harm may scale substantially.

This comparison of the risks profiles of social media platforms, search engines, and ChatGPT poignantly shows that the systemic risks contemplated by the DSA in the context of VLOPs and VLOSEs are present with ChatGPT use as well. If the DSA were considered applicable to ChatGPT, OpenAI would be under an obligation to assess and mitigate such risks. As things stand, however, no such regulatory duty clearly applies, leaving these issues unresolved.

# 4. Conclusion

In this paper we have argued that ChatGPT falls within the scope of the DSA as an online search engine. However, and given it exhibits fundamental characteristics of an online platform, we have further argued that it should be considered as a hybrid of the two types of hosting services, despite it not meeting the public dissemination requirement of online platforms with respect to user prompts.

The paper's main theoretical contribution lies in analyzing and defending the taxonomical position of online search engines within the DSA's intermediary service categories. We conclude that online search engines should, like online platforms, be considered hosting services, an interpretation supported by Botero Arcilia's (2023, pp. 483–485) analysis of how search engines fit within the DSA's broader hosting service framework. We provided systematic textual and legislative analysis demonstrating that search engines store user queries and information at users' request – thereby meeting the hosting service definition – and that the DSA's structure (particularly Art. 24 and Recitals 41, 65, 77) consistently treats search engines alongside platforms as a subcategory of hosting services. We thereby contribute to resolving the regulatory uncertainty surrounding search engines.

Our functional analysis of ChatGPT reveals three distinct pathways to DSA applicability. First, ChatGPT with search unambiguously satisfies the online search engine definition through its real-time web access capabilities. Second, user prompt storage and conversational histories establish a hosting service status through implicit user requests for information retention. Third, custom GPTs explicitly meet online platform criteria by enabling user-generated content storage and public dissemination. Moreover, our comparative assessment of ChatGPT's risk profile suggests this service exhibits systemic risk profiles analogous to traditional VLOSEs and VLOPs across all risk categories specified in Art. 34(2).

This analysis is intended to explore the resilience of the DSA with regard to the increasingly evolving nature of digital services, where the lines between traditional online search engines and online platforms will increasingly blur. In this regard, our proposal anticipates and accommodates this evolution towards systems capable of mediating online activity in new ways. However, several limitations warrant acknowledgment. First, our analysis focuses primarily on ChatGPT as a standalone service, leaving open questions about how the framework applies to generative AI functionalities embedded within existing search engines (such as Google's AI Overviews or Bing's Copilot integration). Second, the lack of publicly available EU-specific user data for custom GPTs and similar features creates empirical gaps that may affect threshold determinations for VLOP designation. Third, the rapid evolution of these technologies means that the features analysed here – particularly ChatGPT's search functionality and custom GPTs – may undergo substantial changes that alter their regulatory classification. Future research should systematically examine how the hybrid classification framework proposed here applies to other generative AI services, including those integrated into traditional platforms and search engines. Additionally, empirical studies tracking user behaviour across these hybrid services would strengthen our understanding of how systemic risks manifest in practice.

Finally, while this paper demonstrates that ChatGPT fits within the DSA's existing definitional framework, deeper questions remain about whether the DSA's safe harbour logic – predicated on intermediary non-liability for third-party content – remains coherent when applied to generative AI systems that produce their own outputs. Whether the AI Act provides a more suitable regulatory model for such systems, or whether hybrid governance approaches are needed, represents an important avenue for future theoretical inquiry (For an assessment of the DSA's "meta-regulation", see Hoboken et al., 2023, pp. 215–219).

Overall, since ChatGPT has crossed the VLOSE user threshold and should be designated as such. Additionally, we propose that the risk assessment and mitigation obligations should also take into account the risks emanating from the online platform characteristics of ChatGPT.